\documentclass[12pt,letterpaper]{article}
\usepackage[margin=1.2in]{geometry}
\usepackage[english]{babel}
\usepackage[utf8x]{inputenc}
\usepackage{amsmath}
\usepackage{amssymb}
\usepackage{mhchem}
\usepackage[retainorgcmds]{IEEEtrantools}
\usepackage{graphicx}
\usepackage{tabularx}
\usepackage{subfig}
\usepackage{kpfonts}    
\usepackage{setspace}
\usepackage{microtype} 
\usepackage{booktabs}   
\usepackage{bm}         
\usepackage{listings}   
\usepackage{verbatim}   
\usepackage{color}  
\usepackage{float}
\usepackage[hidelinks]{hyperref}
\usepackage[colorinlistoftodos]{todonotes}
\usepackage{natbib}
\usepackage{authblk} 
\title{\textbf{The Role of Carbon Pricing in Food Inflation: Evidence from Canadian Provinces}}
\author{Jiansong Xu}
\date{\textit{This draft was completed on May 22, 2024}}
\affil{Department of Resource Economics and Environmental Sociology, University of Alberta}

\begin{document}

\maketitle

\begin{abstract}
    In the search for political-economic tools for greenhouse gas mitigation, carbon pricing, which includes carbon tax and cap-and-trade, is implemented by many governments. However, the inflating food prices in carbon-pricing countries, such as Canada, have led many to believe such policies harm food affordability. This study aims to identify changes in food prices induced by carbon pricing using the case of Canadian provinces. Using the staggered difference-in-difference (DiD) approach, we find an overall deflationary effect of carbon pricing on food prices (measured by monthly provincial food CPI). The average reductions in food CPI compared to before carbon pricing are $2\%$ and $4\%$ within and beyond two years of implementation. We further find that the deflationary effects are partially driven by lower consumption with no significant change via farm input costs. Evidence in this paper suggests no inflationary effect of carbon pricing in Canadian provinces, thus giving no support to the growing voices against carbon pricing policies.
\end{abstract}

\section{Introduction}
Carbon pricing is gaining popularity among countries that are ambitious about the mitigation of greenhouse gases. According to \cite{worldbank_dashboard}, 39 national jurisdictions have carbon pricing in effect as of 2023, covering 23\% (\ce{11.66 Gt CO_2eq}\footnote{\ce{CO_2eq = carbon dioxide equivalent.}}) of global greenhouse gas (GHG) emissions. Some representative countries/regions are Europe, Canada, New Zealand, and Mexico. Two different approaches exist for implementing carbon pricing: carbon tax and emission trade system (ETS, also referred to as cap-and-trade) (\citealt{lse_carbonpricing_2019}). The former is a tax attached to the production or consumption of each unit of pollution or its source. In contrast, the latter allocates tradeable emission permits to producers and facilitates their exchange.

While the influence on food prices has become a common concern for carbon pricing, economists have found little evidence for the inflationary effects (e.g., \citealt{tombe_energy_2023}, \citealt{moessner_effects_2022}). Some studies even see deflationary effects from carbon pricing policies (e.g., \citealt{konradt_carbon_2021}, \citealt{mckibbin_climate_2021}). This counter-intuitive finding may be explained by the tax revenue recycling (\citealt{beck_carbon_2015}) and shifts in consumer preference (\citealt{NGFS_climate_monetary}). Food is considered one of the oil price-sensitive items given its reliance on transportation, heating, and cooling (\citealt{chen_rise_2023}). Therefore, the responsiveness of food prices to carbon pricing---which mainly targets the energy sector---has received special attention in previous studies (\citealt{konradt_carbon_2021}; \citealt{moessner_effects_2022}). However, while providing valuable insights, investigations of the long-run dynamic effects of carbon pricing on food prices are missing in earlier research (with the exception of \citealt{konradt_carbon_2021}). Besides, previous case studies of Canada mainly focus on the province of British Columbia. Carbon pricing in other Canadian jurisdictions could expand the relevance of these results.

Our main goal is to investigate the short-run and long-run effects of carbon pricing on food prices in Canadian provinces\footnote{We do not consider territories due to the complexity of their food markets. Many remote communities rely on few suppliers for expensive food because of the difficulty of shipping food to these locations.}. As a country with almost two decades of experience of carbon pricing, Canada is an exemplary case for study. For this reason, many previous studies on carbon pricing have been case studies of Canada (e.g., \citealt{konradt_carbon_2021}, \citealt{benjamin2022carbon}). To our knowledge, no previous study of the Canadian experience has taken the approach applied here. 

We apply difference-in-difference (DiD) methods to capture the average effects of carbon pricing policies on food CPI for Canadian provinces with effective carbon pricing. Data used in the model are sourced from \cite{statistics_canada_consumer_noyear} and the \cite{bls_us_cpi}. Our results suggest a deflationary effect of carbon pricing on food prices. We also find that such a deflationary effect is more prominent in the long run. In addition, this study provides evidence that food deflation after carbon pricing is mainly attributable to reductions in consumption, while farm production costs are not sensitive to these policies.

\section{Carbon pricing systems in Canada}
The history of carbon pricing in Canada traces back to 2007 when the provinces of Quebec and Alberta implemented the country's first carbon tax and cap-and-trade, respectively (\citealt{statscan_carbonpricing_system}). Qu\'ebec implemented Canada's first emission tax targeting energy producers in October 2007 (\citealt{quebec_tax}). However, the tax rates were too low to induce behavioral changes (\citealt{yamazaki2017jobs}; \citealt{hanoteau2019_qc_tax}
). While Manitoba also implemented an emission tax in January 2012 (\citealt{mb_tax}), it covered only coal and petroleum coke. Therefore, the taxes in Qu\'ebec and Manitoba are not considered in this study.

Canada's first standard carbon pricing policy was implemented in 2007 in Alberta as a cap-and-trade system with production-and-trade options for offsets. Under the Alberta Specified Gas Emitters Regulation (SGER) (\citealt{ab_capandtrade}), only a small portion of industries are required to reduce emission intensity measured by tonnes of \ce{CO_2eq} per unit of output per year. Emitters who are mandated to mitigate have four options: (1) reduce emission intensity themselves, (2) purchase credits/permits from other regulated emitters, (3) purchase in-province emission offsets, the supply of which relies significantly on agricultural and land activities, or (4) pay a carbon levy into a fund managed by the provincial government (\citealt{swallow_2016_ab_offset}).
Agriculture is not regulated under SGER, but it is an important supplier of offsets. This policy became North America's first multi-sector emission trading system that encourages fast-abating producers to take more abatement actions and sell leftover allowances to the slower ones (\citealt{ab_capandtrade}).

The first time the name ``carbon tax" was used in a Canadian legislative document was in British Columbia, where the province initiated a tax on energy purchases in July 2008 (\citealt{bc_tax}). The ineffective tax in Qu\'ebec was replaced by a cap-and-trade system in January 2013 (\citealt{quebec_capandtrade_2018}). At the end of 2016, Alberta, British Columbia, and Qu\'ebec had provincial carbon pricing systems in effect.

In December 2016, Canada announced the Pan-Canadian Framework (PCF) that aims to reduce GHG emissions by \ce{219 Mt CO_2eq} from the 2016 level\footnote{This target was subsequently replaced by \textit{Net-zero emissions by 2025} (\citealt{goC_net-zero_2023}).}. PCF demands all provinces have carbon pricing matching the federal benchmark by 2018. The federal government inserts a ``backstop" to any shortage in carbon pricing to help all provinces meet the benchmark (\citealt{winter_carbon_2020}).

After the deadline for carbon pricing set by the PCF, all Canadian provinces and territories have carbon pricing systems in place, with federal backstops applied partially or in full. Some existing carbon pricing systems before the PCF were also replaced by new systems. A detailed description of post-PCF carbon pricing in each province can be found in \cite{auditor_carbon_2022} and \cite{winter_carbon_2020}.

\section{Related literature}
Although the literature on the economic effects of carbon pricing is not rich, studies generally disagree with the opinion that carbon pricing causes affordability issues. For example, \cite{tombe_energy_2023} examine the effective carbon tax rate in British Columbia and find that carbon tax only increased the prices of most items by less than $0.3\%$, with the effect on food being slightly higher ($0.33\%$). \cite{moessner_effects_2022} find similar results among the $35$ OECD countries. They suggest that by increasing the carbon price by $\$10$ per ton \ce{CO_2eq}, food and core CPI\footnote{Consumer price index} show no significant response. \cite{kanzig_unequal_2023} separately investigates headline and core CPI\footnote{Core CPI excludes food and energy} and finds significant increases in both indices, but the overall effect is minimal compared to the increase in energy price.

Some studies find adverse impacts on inflation. \cite{mckibbin_climate_2021} show that while a carbon tax has a small positive impact on the headline inflation of Europe, the core inflation indeed decreases because of carbon tax. This disparity provides evidence that food and energy may respond to carbon pricing differently compared to other commodities. Another study by \cite{konradt_carbon_2021} suggests significant but small deflation effects from carbon taxes in Canada and Europe.

The counter-intuitive effects of carbon pricing on inflation are attributable to several factors. First, consumers may switch to green products due to the policy (\citealt{NGFS_climate_monetary}). This substitution behavior will, to some extent, offset the effects on the cost of living. Second, the presence of an output-based pricing system (OBPS) for large emitters, such as in Alberta, may relieve the increases in household expenditures caused by carbon pricing. \cite{winter_carbon_2023} find that an OBPS mitigates the effects of carbon pricing on electricity costs for provinces with emission-intensive electricity grid, through which it reduces the costs embedded in the production of goods and services. Third, results of \cite{konradt_carbon_2021} suggest that the increase in energy price due to carbon tax is more than offset by the decrease in prices of services and other tradables. This phenomenon may be explained by the falling income of higher-income households, cheaper real estate, and energy-intensive durables (\citealt{konradt_carbon_2021}). Another potential reason is the recycling of proceeds. \cite{beck_carbon_2015} find that the carbon tax rebate in British Columbia increases the real income of lower-income families. Lastly, \cite{andersson_climate_2020} suggest that carbon pricing encourages innovations in renewable energy. However, the speed of innovation makes renewable energy unlikely to explain the price changes in the short run. 

To assess mechanisms of changes in food prices, we focus on the demand and supply shocks associated with carbon pricing. Specifically, we access changes in consumption per capita and farm input costs after carbon pricing.

\section{Empirical analysis}
\subsection{Data}
We use the monthly food CPI of Canadian provinces between January 2005 and December 2016 (inclusive) from \cite{statistics_canada_consumer_noyear}. The original data starts from 2000 and is constantly being updated. We consider only the window between 2005 and 2016 for two reasons. First, because the data is monthly, there are too many pre-treatment periods before 2005 which may capture the effects of irrelevant events. Second, the PCF, which started in 2017, led to nationwide carbon pricing and updated some existing provincial policies. Identifying the effect of PCF requires more complex designs and does not fall within the scope of this study. For this reason, we leave carbon pricing policies after 2016 for future research.

The final sample data we use includes ten provinces: Alberta (AB), British Columbia (BC), Qu\'ebec (QC), Manitoba (MB), Saskatchewan (SK), Ontario (ON), New Brunswick (NB), Newfoundland and Labrador (NL), Nova Scotia (NS), and Prince Edward Island (PEI). The provinces of AB, BC, and QC are labeled as the ``treated provinces" or ``treated group" as they had effective carbon pricing policies during the study period. Other provinces are labeled as the ``controlled provinces" or ``controlled group". In addition, the start month of carbon pricing in each treated province is marked, which are
\begin{enumerate}
    \item AB: 2007-07
    \item BC: 2008-07
    \item QC: 2013-01
    \item Other provinces have no start months\footnote{We place ``0" in each entry of the start month for controlled provinces. However, this value is arbitrary as they will be canceled out later on in the staggered difference-in-difference model.}.
\end{enumerate}

The monthly food CPI data for ten provinces and twelve years provides a total of $1440$ observations.

\subsection{Difference-in-difference (DiD)}
We first use a simple DiD method to identify the average treatment effect on the treated (ATT) of carbon pricing on log food CPI (base $= e$) in each treated province. The empirical model for each treated province is specified as

\begin{equation}
    \label{eq:did}
    LFCPI_{i,t} = \tau CP_{i,t} + \delta_{i} + \mu_{t} + \epsilon_{i,t},
\end{equation}

where $i$ and $t$ denote the group membership (a treated province or the controlled provinces) and month. $LFCPI_{i,t}$ is the logarithmic monthly food CPI in month $t$ and province $i$. $CP_{i,t}$ is a dummy variable that takes on the value of $1$ when an observation belongs to the treated provinces and is observed after the implementation of carbon pricing, which is equivalent to the interaction of a province dummy and a month dummy. $\tau$ is the average treatment effect of carbon pricing on log food CPI of the treatment group (i.e., ATT). Parameters $\delta_{i}$ and $\mu_{t}$ are province and month fixed effects. Lastly, $\epsilon_{i,t}$ is the disturbance term. Note that our specification resembles the robustness check section of \cite{konradt_carbon_2021}. However, this model extends to more provinces and focuses on food commodities.

We estimate \autoref{eq:did} for each treated province (Alberta, Quebec, British Columbia) against the whole control group. The food CPI for the latter is taken as the group average. This repeated pairwise DiD allows us to assess the ATT in each treated province.

We also adopt a staggered DiD design to accommodate the different treatment timings across provinces and capture the long-run treatment effects. Following \cite{callaway2021difference}, the staggered DiD model in our study is

\begin{equation}
    \label{eq:strgdid}
    LFCPI_{i,t} =  \Sigma^{-2}_{e=-K} \tau^{pre}_e D^{e}_{i,t} + \Sigma^{L}_{e=0} \tau^{post}_{e} D^{e}_{i,t} + \delta_{i} + \mu_{t} + \upsilon_{i,t},
\end{equation}

where $e$ denotes each lag or lead to the treatment. $K$ and $L$ are positive constants denoting the first lead and the last lag observed. $D^{e}_{i,t} = 1\{t - t^{treat}_i = e\}$ represents the interaction between treated provinces and treated months. Coefficients $\tau^{pre}_e
$ and $\tau^{post}_{e}$ are effects corresponding to each period before or after the treatment (i.e., leads and lags). All coefficients are normalized with respect to the coefficient of period $e=-1$. We are interested in the treatment effects averaged across treated provinces in each post-treatment period ($\tau^{post}_{e}$), which is also known as the group-time average treatment effects. Fixed effects are again denoted by $\delta_{i}$ and $\mu_{t}$, and $\upsilon_{i,t}$ is the new error term. Because of the inclusion of multiple time periods and treated groups, staggered DiD is also known as a dynamic two-way fixed effects or event study.  

Another benefit of staggered DiD is that the model specification is itself a falsification test. We can learn the validity of the treatment from the difference between pre- and post-treatment effects. We should observe insignificant pre-treatment effects ($\tau^{pre}_e$) and significant post-treatment effects ($\tau^{post}_{e}$) if changes in the output variable are caused by the treatment. In addition, the post-treatment coefficients show the evolution of treatment effects in the long run. A similar approach is used in \cite{autor2003outsourcing}.  

\begin{figure}[!htpb]
    \centering
    \includegraphics[width = 14cm, height = 10cm]{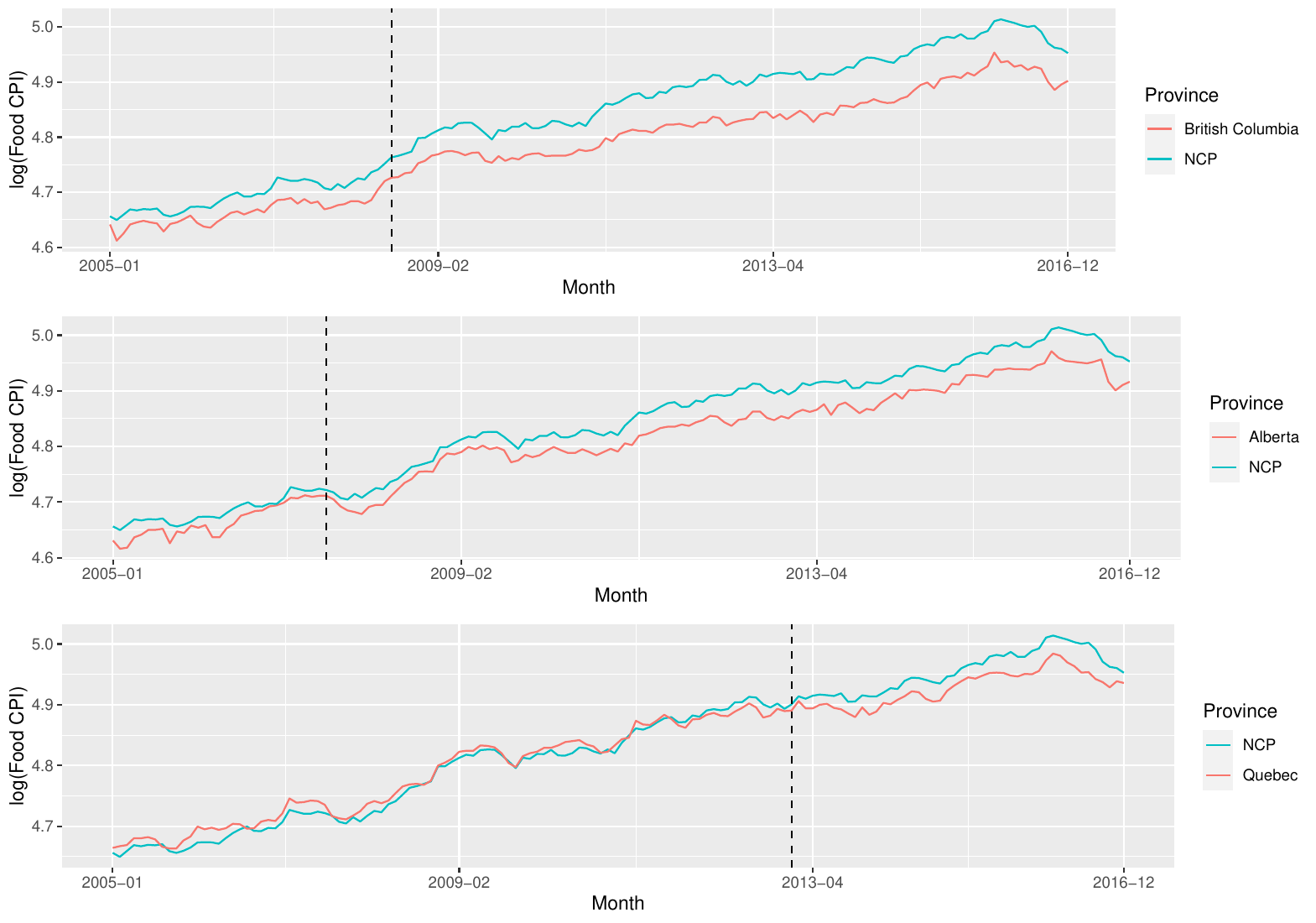}
    \caption{Monthly logarithmic food CPI (not seasonally adjusted) between 2000 and 2023. Source: author's own calculation with data from \cite{statistics_canada_consumer_noyear}}
    \label{fig:provincial_cpi}
\end{figure}

We illustrate the trends of food inflation (log(food CPI)) in treated and controlled groups in Figure \ref{fig:provincial_cpi}. NCP (no carbon pricing) represents the average of the control group. Each vertical dashed line indicates the month when a provincial carbon pricing policy was enacted. Before each treated province receives treatment, food inflation in treated and controlled provinces follows roughly the same increasing pattern. However, the growth of food inflation in each treated province starts to slow down after treatment, which is reflected by the expanding gaps between the red and blue curves in \autoref{fig:provincial_cpi}. The small divergence between the trends of Quebec and controlled provinces before treatment is unlikely to compromise the validity of the treatment due to its size. Such short and small pre-treatment changes could be the results of the anticipation of policies. We rely on the significance of the pre-treatment effects in \autoref{eq:strgdid} to test the parallel trend assumption. Additionally, the trends of Quebec and controlled provinces overlap until the treatment (cap-and-trade) is close, which agrees with \cite{hanoteau2019_qc_tax} that the carbon tax in Quebec was ineffective.

Manitoba is included in the controlled group due to the limited coverage of its fuel tax. Nevertheless, the literature provides insufficient evidence regarding the effectiveness of this tax, rendering the group identity of Manitoba uncertain. To address the potential model misspecification, we conduct a falsification test by moving Manitoba from the controlled group to the treated group in another staggered DiD regression.

\section{Regression results}
We first run separate DiD models for each pairwise comparison between one treated province and the controlled provinces. From the results presented in \autoref{tab:did_pair}, all models suggest negative effects of carbon pricing on food CPI. Each column of \autoref{tab:did_pair} contains results for one treatment province, as well as the type of carbon pricing applied (carbon tax or cap-and-trade). BC has the largest (negative) estimated effect, stating that the carbon tax program in BC is linked to a reduction in food CPI by $4.792$. Other provinces also show significant negative effects. The effect in Alberta is only half of other provinces. The overall effect of carbon pricing on food CPI based on separate pairwise DiD models lies between $-2.5$ and $-5$, which confirms the results by \cite{mckibbin_climate_2021} and \cite{konradt_carbon_2021} that carbon pricing is deflationary. 

\begin{table}[!htbp] \centering 
\small
  \caption{Regression results of non-staggered DiD} 
  \label{tab:did_pair} 
\begin{tabular}{@{\extracolsep{5pt}}lccc} 
\\[-1.8ex]\hline 
\hline \\[-1.8ex] 
 & \multicolumn{3}{c}{\textit{Dependent variable:}} \\ 
\cline{2-4} 
\\[-1.8ex] & \multicolumn{3}{c}{Food CPI} \\ 
 & BC: tax & AB: trade & QC: trade \\ 
\hline \\[-1.8ex] 
 ATT & $-$0.032$^{***}$ & $-$0.016$^{***}$ & $-$0.031$^{***}$ \\ 
  & (0.002) & (0.002) & (0.002) \\ 
  & & & \\ 
Province FE & Yes & Yes & Yes \\ 
Month FE & Yes & Yes & Yes \\ 
\hline \\[-1.8ex] 
Observations & 266 & 266 & 266 \\ 
R$^{2}$ & 0.631 & 0.274 & 0.680 \\ 
Adjusted R$^{2}$ & 0.254 & $-$0.468 & 0.352 \\ 
F Statistic (df = 1; 131) & 224.444$^{***}$ & 49.473$^{***}$ & 277.900$^{***}$ \\ 
\hline 
\hline \\[-1.8ex] 
\textit{Note:}  & \multicolumn{3}{r}{$^{*}$p$<$0.1; $^{**}$p$<$0.05; $^{***}$p$<$0.01} \\ 
\end{tabular} 
\end{table} 

\begin{figure}[!hbpt]
    \centering
    \includegraphics[width = 13cm, height = 9cm]{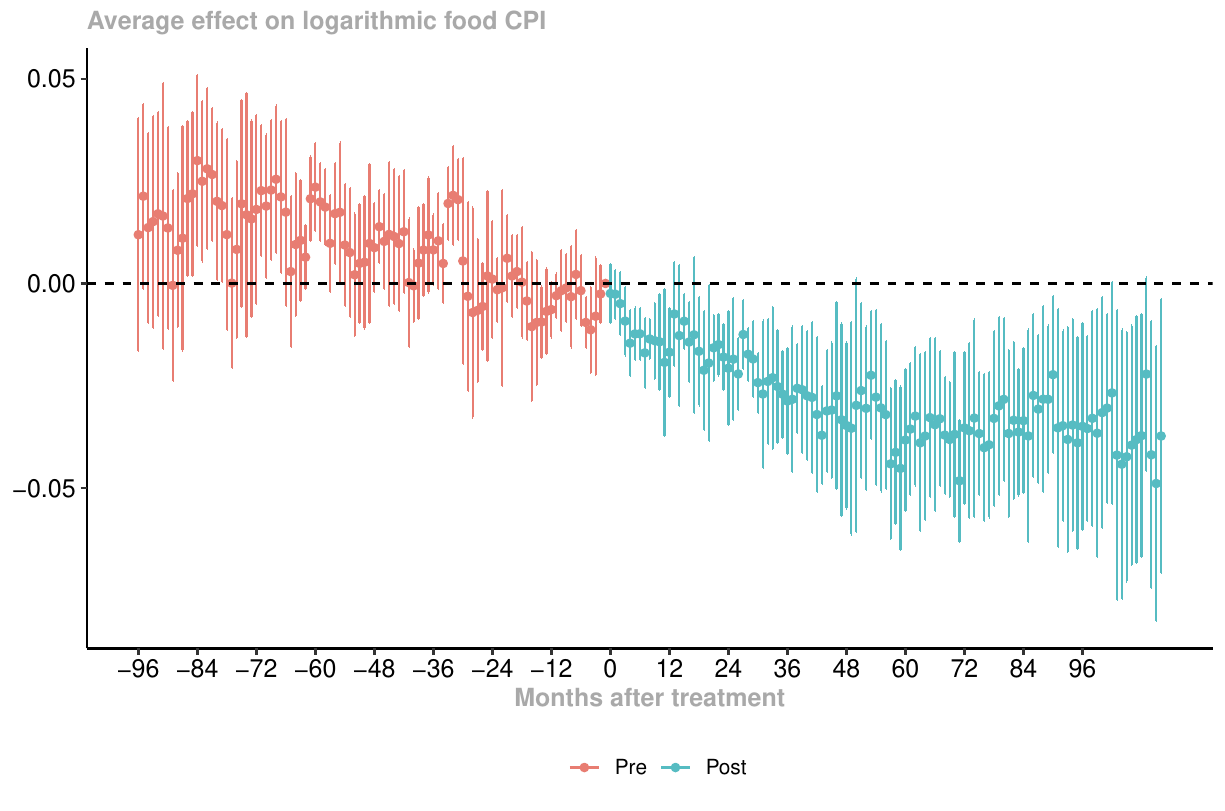}
    \caption{Group-time average treatment effect by length of exposure to the treatment}
    \label{fig:did_staggered}
\end{figure}

When the DiD model has a staggered design as in \autoref{eq:strgdid}, we observe some interesting patterns in long-term effects. The estimated average treatment effects and their confidence intervals are plotted in \autoref{fig:did_staggered}. The horizontal axis contains the leads (positive values) and lags (negative values) to the treatment. The first observation is that the effects of carbon pricing on food CPI are also deflationary as in separate pairwise models. The effects appear to vary over time. The immediate effects after carbon pricing are either weak or insignificant, while they become more prominent after 24 months (2 years) of the treatment. The difference between the short-run and long-run estimates suggests that the food deflation associated with carbon pricing requires some ``rendering-in" time, which could be the time spent on the adjustment of variable and quasi-variable assets, as well as consumption patterns.

In summary, the cost of food on average decreases by $\approx 2\%$ and $\approx 4\%$ within and after two years of carbon pricing, relative to the pre-treatment periods\footnote{We are able to interpret the effects relative to pre-treatment periods because the pre-treatment effects are mostly insignificant.}.

Our next finding relates to the parallel trend assumption that is fundamental for DiD studies. Most of the confidence intervals of pre-treatment effects cover 0, especially those in the last 36 months (3 years) before treatment. Therefore, despite some small splits in pre-treatment trends shown in \autoref{fig:provincial_cpi}, we do not reject the parallel trend assumption.

The last observation, or the problem, is that the confidence intervals are wide. This issue arises from the macro nature of our data---there are at most $10$ observations (provinces) in each period. While the estimated effects are still informative, we refrain from testing any hypotheses other than the confidence intervals as the test statistics will lack reliability. On the other hand, despite the uncertainty between the upper and lower bounds of estimates, the overall treatment effects still show clear reductions in food CPI after carbon pricing.

\begin{figure}[!hbpt]
    \centering
    \includegraphics[width = 12.5cm, height = 11cm]{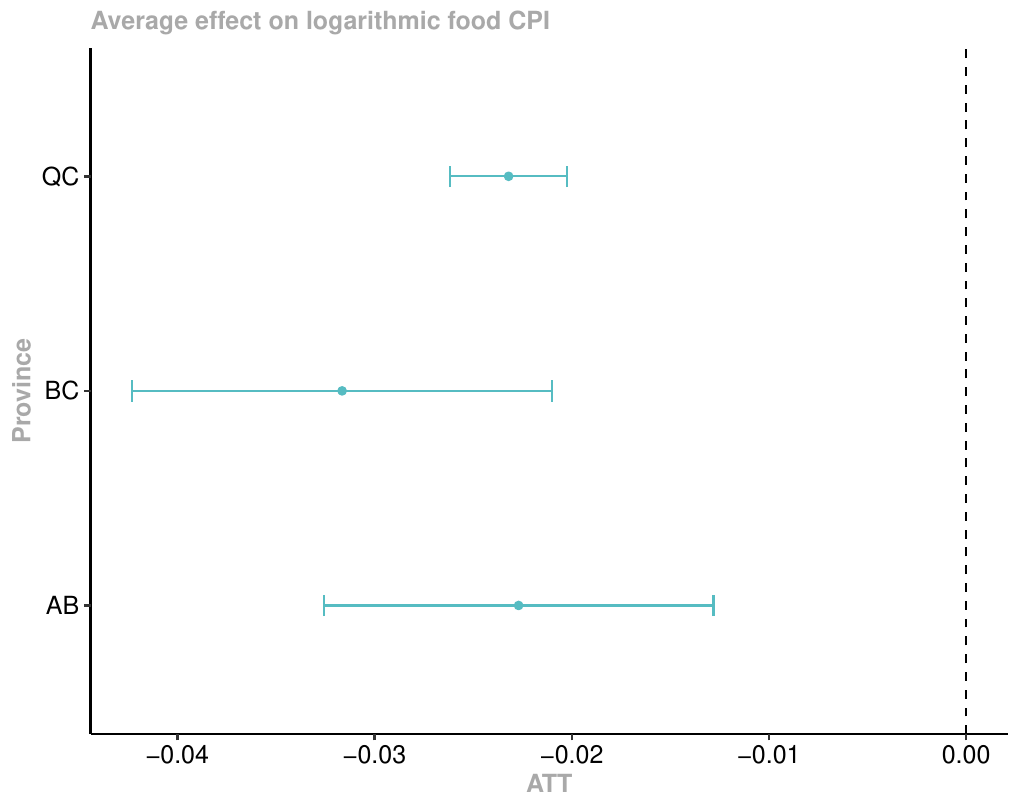}
    \caption{Treatment effect by province}
    \label{fig:did_staggered_group}
\end{figure}

We also estimate the province-level treatment effects to examine provincial heterogeneity (\autoref{fig:did_staggered_group}). All treated provinces experienced significant decreases in food CPI after carbon pricing. The food CPI of British Columbia reduced by the most ($\approx -3\%$). The effects in the other two provinces are around $-2.2\%$. Nonetheless, none of provincial effects suggest drastic changes in food price.

In summary, the effects of carbon pricing on food CPI are overall deflationary. The decreases in the cost of food were minimal/insignificant in the first two years of carbon pricing but gradually intensified and plateaued at $\approx -4\%$ relative to pre-treatment. There is a slight upward trend in the effects after seven years of carbon pricing, but the magnitude is incomparable with the earlier negative effects. Hence, at least within an 8-year window (96 months), our results suggest an association between food deflation and carbon pricing policies.

\begin{figure}[!hbpt]
    \centering
    \includegraphics[width = 14cm, height = 8.5cm]{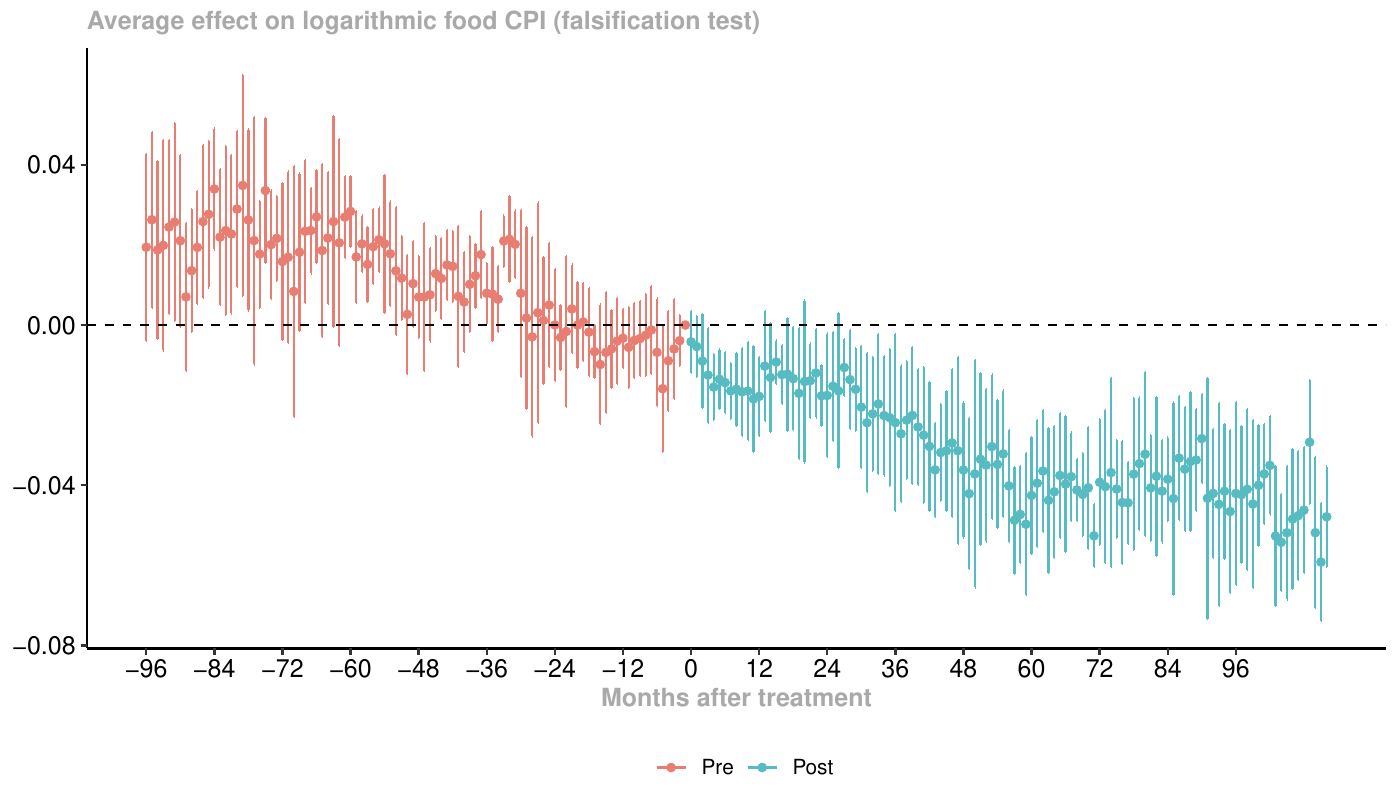}
    \caption{Group-time average treatment effect by length of exposure to the treatment (Manitoba as treated)}
    \label{fig:did_staggered_robtest}
\end{figure}

Lastly, we perform falsification test by re-estimating \autoref{eq:strgdid} with Manitoba defined as a treated province. The post-treatment effects do not change significantly after we redefine the coverage of treatment (see \autoref{fig:did_staggered_robtest}). On the other hand, most pre-treatment effects are significantly different from 0. Although the pre-treatment effects are still stabilized around 0 in the last 36 months before treatment, the parallel trend assumption is less likely to hold when Manitoba is considered as treated instead of controlled.

\section{Mechanism of price adjustment}
So far, our results suggest that food inflation in Canada has indeed decreased because of carbon pricing policies, but the channels of the deflationary effects remain unclear. In this section, we investigate the market-based channels---in other words, the effects of demand and supply shocks.

\begin{figure}[H]
    \centering
    \includegraphics[width = 14cm, height = 10cm]{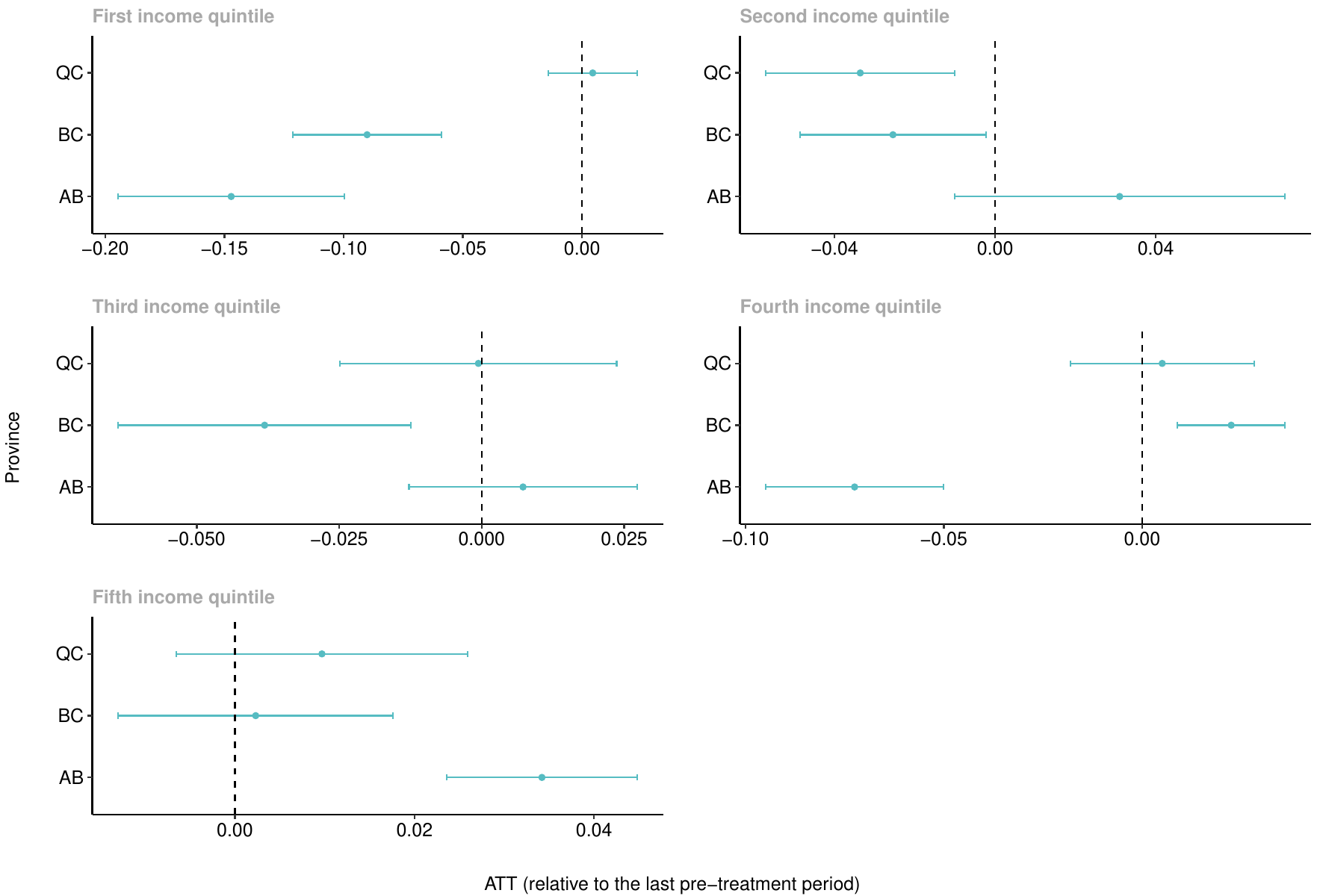}
    \caption{Consumption effect after carbon pricing by income quintile}
    \label{fig:did_consump}
\end{figure}

We first examine the demand-side shock in terms of consumption. We use another staggered DiD model with the same independent variables and log consumption per capita as the new output variable\footnote{Data of consumption is collected by \cite{statcan_distributions}}. Due to limited data availability, we use aggregate consumption instead of food alone. A separate regression is used for each of the five income quintiles of each treated province. The results are plotted in \autoref{fig:did_consump}.

The consumption effects of carbon pricing are heterogeneous across provinces and income quintiles. The first three income quintiles (mid to low income) have clear drops in consumption. BC has reduced consumption in all of the first three income quintiles. The average effects range from $\approx -9\%$ in the first quintile to $\approx -0.035\%$ in the third quintile. Alberta experiences the strongest shrinkage in consumption in the first quintile ($\approx -14\%$), while the effects in the next two quintiles are insignificant. The consumption per capita of Quebec is affected the least, with the most substantial expected reduction being only $\approx -3\%$ in the second quintile. 

The consumption effects are more puzzling in the higher-income quintiles (last two plots of \autoref{fig:did_consump}). Alberta has an average reduction in consumption of $\approx -7\%$ in the fourth quintile, while the average effect was $\approx 3.5\%$ in the highest quintile. It is unclear why the directions of effects differ. However, the overall consumption effect in higher-income quintiles is still negative since the increase in the fifth quintile is small. The other two provinces in upper-income quintiles mostly have insignificant changes in consumption, with BC having a small and significant increase in the fourth quintile. Despite the heterogeneity across income quintiles and provinces, the overall effect of carbon pricing on consumption is predominantly negative. 

Nevertheless, the confidence intervals for the consumption effects are rather wide for many provinces and quintiles, and the mean estimates of Alberta are too large to be credible. Therefore, it is important to regard these estimates with caution. One is advised only to infer the positivity and negativity of the consumption effects from \autoref{fig:did_consump}.

\begin{figure}[!hbpt]
    \centering
    \includegraphics[width = 11cm, height = 9cm]{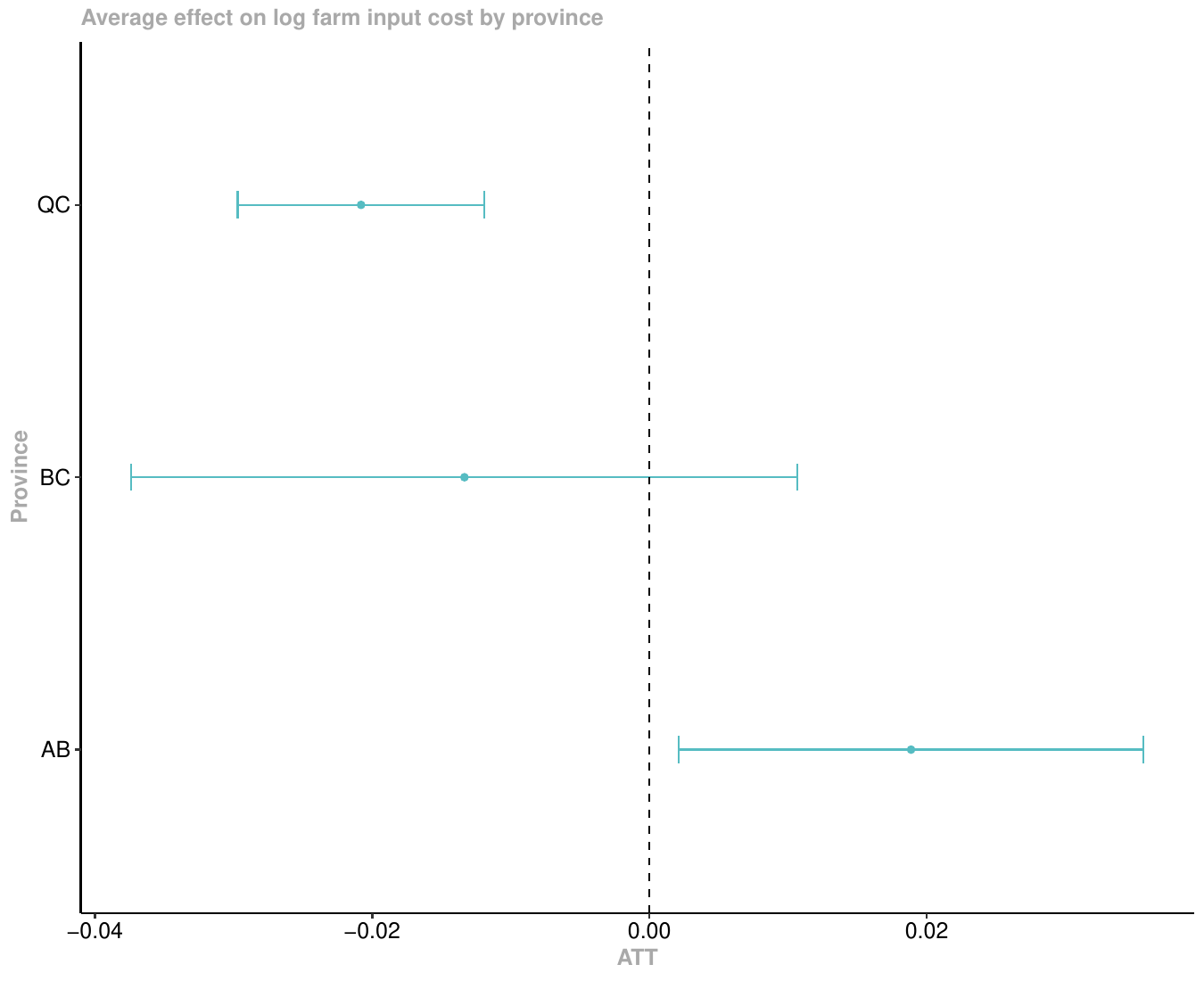}
    \caption{Farm production cost effects after carbon pricing by province}
    \label{fig:did_farmcost_group}
\end{figure}

The supply side effect, on the other hand, does not seem significant. We use staggered DiD once more to regress on the log-transformed quarterly farm input cost index (index for 2002 $=$ 100 before transformation)\footnote{Data of farm input cost index is collected by \cite{statcan_farmcost}}. The results by province are shown in \autoref{fig:did_farmcost_group}. Changes in farm input costs are minimal. Only the farm input cost of Alberta increased by $2\%$ after carbon pricing, while the effects in other treated provinces are either negative or insignificant.

\begin{figure}[!hpbt]
    \centering
    \includegraphics[width = 12cm, height = 9cm]{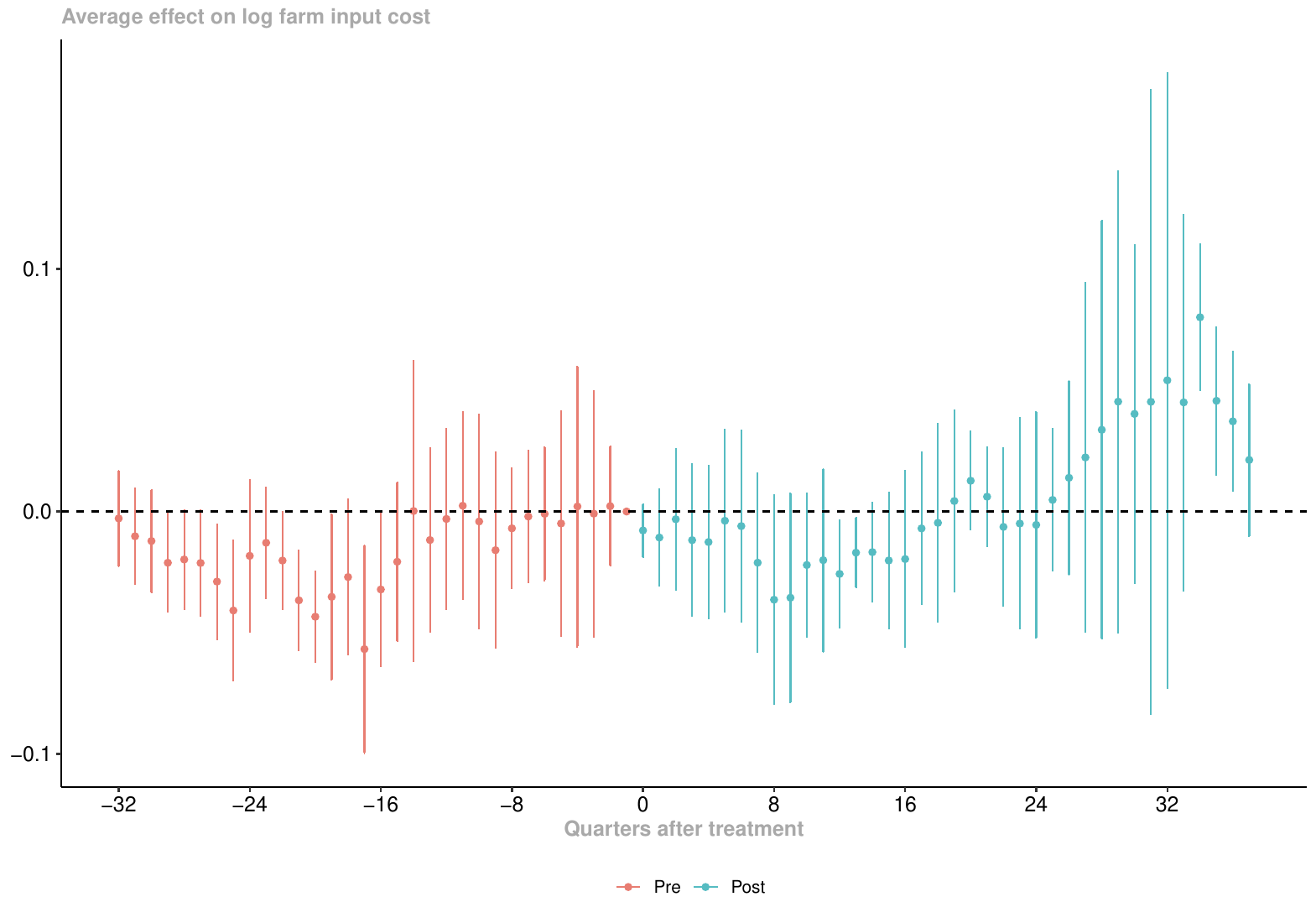}
    \caption{Average effect on farm input cost by length of exposure to the treatment}
    \label{fig:did_farmcost_dynamic}
\end{figure}

We plot the average treatment effects on log farm input cost in \autoref{fig:did_farmcost_dynamic}. Similar to the provincial-level effects shown in \autoref{fig:did_farmcost_group}, there is no significant change in farm input cost after carbon pricing. The average treatment effect has a small bump after 24 quarters (6 years) of treatment. Nevertheless, given that the bumped effects are still insignificant and that 24 quarters are rather far away, we do not consider it an effect of carbon pricing.

The insignificant effect on farm input costs is likely due to the design of carbon pricing policies. In Alberta, the cap-and-trade system encourages farms to reduce emissions and sell their leftover permits to larger emitters. In the province's 2009 summary of SGER (\citeyear{ab_capandtrade}), reduced tillage offset \ce{1,607} thousand tonnes of \ce{CO_2eq} in the 2009 compliance year, making it the largest supplying source of emission offsets in the market. The revenue from offsets may have made up for farmers' loss from abatement actions. In BC, the carbon tax is exempted for fuels used for farming purposes. Although some subsequent production stages are still taxed (e.g., processing, transportation, etc), farm operation takes a significant portion of the total carbon footprint of food. Lastly, the agri-food establishments in Quebec receive free allowances in the cap-and-trade system. Though the direct impact of free allowances on farm production is unclear, one can expect that this policy, to some extent, offsets the cost burden of farmers.

Therefore, we find a significant reduction in consumption (demand side) and an insignificant change in farm input cost (supply side). These analyses suggest that food deflation in treated Canadian provinces is mainly due to reduced demand. It should be noted that this deflation is not necessarily beneficial for consumers. Indeed, the deflated food price is a result of lower consumption rather than a signal of future welfare improvement.

Our results on the consumption effect are somehow not in line with studies focusing on tax revenue recycling. \cite{beck_carbon_2015} and \cite{winter_carbon_2023} both find that carbon pricing harms low-income households less. Contrarily, based on our analysis shown in \autoref{fig:did_consump}, lower-income quintiles are not hurt any less than the higher-income ones. The disagreement between results may be attributable to the methods used. \cite{beck_carbon_2015} and \cite{winter_carbon_2023} use a CGE model and a simulation model with synthetic microdata, respectively, while we use econometric methods to identify causal effects. More research is needed to detail the connection between carbon pricing and consumption.

\section{Conclusion}
This study finds that carbon pricing policies in Canadian provinces have a deflationary effect on food prices. Such an effect requires a ``rendering-in" period of about 24 months (2 years). The actual magnitude of the deflation, however, is still minimal even at its peak. Therefore, carbon pricing should not be blamed for the current high food prices. As pointed out by \cite{tombe_energy_2023}, carbon pricing is not the reason for the high cost of living as its effects on prices are negligible for most commodities\footnote{\cite{tombe_energy_2023} find a negligible increase in food price, which is different from the small negative effect found in this study. Although disagreement exists in the direction of effects, both their and this study conclude that carbon pricing does not drive up inflation}. While the scope of this paper does not include investigating the actual sources of food inflation, recent work by \cite{chen_rise_2023} suggests that the high inflation in Canada is mainly driven by supply shocks in global energy and food markets. 

We find evidence in support of the demand-driven mechanism of deflation, such that consumption per capita decreased due to carbon pricing. The supply-side effect, in contrast, does not contribute much to the deflation. Despite the innocence of carbon pricing regarding food inflation, this finding does not render these policies harmless to general consumers. The decreased food prices are more likely the result of lower consumption rather than a forecast of future welfare gains. Our results on the consumption effects emphasize the necessity of the current federal rebate program targeting lower-income households in Canada. 

Besides consumption, the special treatment farmers receive in light of carbon pricing policies also contributes to lower food prices. Farmers and agri-food producers receive various types of exemptions and aids in provinces that price carbon. These benefits relieve the pressure that carbon pricing could have put on food prices. These exemptions, however, are set to be removed with the amendment and passing of Bill C-234 in 2023. The passed bill made two changes from its original version with respect to farm exemptions: (1) repealed the exemption for fuels used for heating\&cooling of farm structures (e.g., barns), and (2) shortened the sunset clause for propane and natural gas used for grain drying from eight to three years (\citealt{pbo_2024_bill234}). Without additional supportive programs, this shock is likely to pass through the production chain and eventually hit food prices.

Obviously, affecting the affordability of food or other commodities is not the primary target of carbon pricing. The social cost of carbon pricing needs to be justified by the reduction in fossil fuel use and the transition to greener energy sources. The effectiveness of carbon pricing on environmental quality and resource use is a subject for future research. 

A limitation of this study is the level of observation of the data. Models and results capture only variations at the province-month level. One potential solution is to calculate price indices for different municipalities within each province using scanner data.

\clearpage
\bibliographystyle{chicago}
\bibliography{main.bib}
\end{document}